\documentstyle{elsart}
\begin{document}

\newcommand{\dd}{\mbox{d}}
\newcommand{\vecc}[1]{{\bf #1}}

\begin{frontmatter}

\title{ Small--Angle Bhabha Scattering at LEP1. \\
        Analytical Results  for Wide--Narrow Angular Acceptance. }

\author[LTPH]{A.B. Arbuzov},
\author[KFTI]{G.I. Gach},
\author[KFTI]{V.Yu. Gontchar},
\author[LTPH]{E.A. Kuraev},
\author[KFTI]{N.P. Merenkov} and
\author[Parma]{L. Trentadue}

\address[LTPH]{Bogoliubov Laboratory of Theoretical Physics, JINR,
Dubna, 141980, Russia}
\address[KFTI]{Institute of Physics and Technology,
Kharkov, 310108, Ukraine}
\address[Parma]{Dipartimento di Fisica, Universit\'a  di Parma and INFN,
Gruppo Collegato di Parma, 43100 Parma, Italy}

\begin{abstract}
Analytical formulae for small--angle Bhabha scattering
cross--section  at LEP1 are given for the case of wide--narrow angular acceptance.
Inclusive and calorimeter event selections are considered. Numerical
results are presented. \\[.4cm]
\noindent
PACS~ 12.20.--m Quantum electrodynamics, 12.20.Ds Specific calculations
\\ \noindent
\end{abstract}

\end{frontmatter}

\section*{Introduction}

    The small--angle Bhabha scattering (SABS) process is used to measure
luminosity of electron--positron colliders. At LEP1 an experimental accuracy
on the luminosity better then one per mille has been reached~[1].
To estimate the total accuracy a systematic theoretical error must be
added. That is why in recent years a considerable attention has been
devoted to theoretical investigation of SABS cross--section~[2--11].

The theoretical calculation of SABS cross--section at LEP1
has to cope with two problems. The first one is the description
of experimental restrictions in terms of final particles phase
space  used for event selection. The second one concerns
the computation of matrix element squared with the required
accuracy. There are two approaches for the theoretical study
of SABS at LEP1: the one basing on Monte Carlo (MC)
programs~[2--6] and the other using semi--analytical
calculations~[7--11].

The advantage of MC method is the possibility to model different
types of detectors and event selections~[2]. But at this approach
some problems with exact matrix element squared exist. Contrary,
the advantage of the analytical approach is the possibility to use
exact matrix element squared and its defect is a low mobility
relative the change of an experimental conditions for event
selection. Nevertheless, the analytical calculations are
of great importance because they allow to check numerous MC
calculations for different types of {\em ideal\/} detectors.

In this letter we list some analytical results for SABS
cross--section at LEP1 suitable for inclusive (labeled in~[2]
as  BARE1) and calorimeter (CALO1 and CALO2) event selections
in the case of asymmetrical wide--narrow circular detectors.
We give analytical formulae for the full first order correction
to the cross--section as well as for leading second and third
order ones. Our numerical estimations include also
next--to--leading second order contribution in the case BARE1.

\section{First order corrections}

Let us consider at first BARE1 event selection. We introduce
the dimensionless quantity
$\Sigma = Q_1^2\sigma_{\mathrm{obs}}/(4\pi\alpha^2),$
where $Q_1^2 = \varepsilon^2\theta_1^2$ ($\varepsilon $ is
the beam energy and $\theta_1$ is the minimal scattering angle
for wide circular detector) and $\sigma_{\mathrm{obs}}$ is an
{\em experimentally\/} observable (by means of {\em ideal\/}
detectors) cross--section. Then the first order QED correction
can be written as follows:
\begin{eqnarray}
\Sigma_1 &=& \frac{\alpha}{2\pi}\Biggl\{
\int\limits_{1}^{\rho_3^2}\frac{\dd z}
{z^2}\int\limits_{x_c}^{1}\biggl[\biggl(-\frac{1}{2}\delta(1-x)
+ (L-1)P^{(1)}(x)\biggr)(\Delta_{42}+\Delta_{42}^{(x)})
+ \frac{1+x^2}{1-x}\widetilde K \biggr]\dd x  \nonumber \\ \label{1}
&+& \int\limits_{\rho_2^2}^{\rho_4^2}\frac{\dd z}{z^2}
\int\limits_{x_c}^{1}\biggl[\biggl(-\frac{1}{2}\delta(1-x)
+ (L-1)P^{(1)}(x)\biggr) (1 + \Theta_3^{(x)})
+ \frac{1+x^2}{1-x}K\biggr]\dd x\Biggr\} \, ,
\end{eqnarray}
where
\begin{eqnarray}
\widetilde K(x,z;\rho_4,\rho_2) &=& \frac{(1-x)^2}{1+x^2}(\Delta_{42}
+ \Delta_{42}^{(x)}) + \Delta_{42}
\widetilde L_1 + \Delta_{42}^{(x)}\widetilde L_2
+ (\overline\Theta_4^{(x)} - \Theta_2^{(x)})\widetilde L_3  \nonumber \\
&+& (\overline\Theta_4 - \Theta_2)\widetilde L_4 \, ,
\nonumber \\ \label{2}
K(x,z;\rho_3,1) &=&
\frac{(1-x)^2}{1+x^2}(1+\Theta_3^{(x)}) + L_1 + \Theta_3^{(x)}L_2
+ \overline\Theta_3^{(x)}L_3 \, ,
\end{eqnarray}
and $L=\ln(zQ_1^2/m_e^2)$ is the {\em large logarithm\/};
parameter $x_c$ puts the restriction on the energies of
the final electron and positron:
$\varepsilon_1\varepsilon_2 \geq x_c\varepsilon^2$.
In Eqs.(1) and (2) we used the following notation for
$\Theta$--functions and $\widetilde L_i$:
\begin{eqnarray*}
\Theta_{i}^{(x)} &=& \Theta(x^2\rho_{i}^2-z),\quad
\Theta_{i} = \Theta(\rho_{i}^2
-z),\quad  \overline\Theta{i}^{(x)} = 1 - \Theta_{i}^{(x)},\\
\overline\Theta_{i} &=& 1 - \Theta_{i}\, , \quad
\Delta_{42} = \Theta_4 - \Theta_2, \quad
\Delta_{42}^{(x)} = \Theta_4^{(x)} - \Theta_2^{(x)}, \\
\widetilde L_1 &=& \ln\left|
\frac{(z-\rho_2^2)(\rho_4^2-z)x^2}{(x\rho_4^2-z)
(x\rho_2^2-z)}\right|,\ \
\widetilde L_2 =\ln\left|\frac{(z-x^2\rho_2^2)(x^2\rho_4^2-z)}
{x^2(x\rho_4^2-z)(x\rho_2^2-z)}\right| , \\
\widetilde L_3 &=& \ln\left|
\frac{(z-x^2\rho_2^2)(x\rho_4^2-z)}{(x^2\rho_4^2-z)
(x\rho_2^2-z)}\right|,\ \ \ \widetilde
L_4 =\ln\left|\frac{(z-\rho_2^2)(x\rho_4^2-z)}
{(\rho_4^2-z)(x\rho_2^2-z)}\right| .
\end{eqnarray*}
The quantities $L_i$ can be obtained from $\widetilde L_i$ by
the substitution $\rho_4 \rightarrow \rho_3, \ \rho_2 \rightarrow 1$.
Here we use the same notations as in~[10]. Note only that for
wide--narrow angular acceptance
\begin{eqnarray}
\rho_3 > \rho_4 > \rho_2 > \rho_1 = 1 \ ,\qquad
\rho_i = \frac{\theta_i}{\theta_1} \, ,
\end{eqnarray}
where $\theta_i$ are the limiting angles of the circular
detectors (see Section~3). Function $P^{(1)}(x)$ defines
the iterative form of the non--singlet electron structure
function (see for example~[6]). The first (second) line in the
r.h.s. of Eq.(1) is the contribution due to photon emission by
positron (electron). The terms accompanied with $x$--dependent
$\Theta$--functions under the integral sign correspond to
the initial--state correction while the rest belongs to
the final--state one.

The CALO1 cluster is the cone with
angular radius $\delta = 0.01$ around the direction of the final
electron (or positron) momentum. If photon belongs to the cluster
the whole cluster energy is measured, and electron may have any
possible energy. Therefore, the limits of $x$--integration for
$\Sigma_{\mathrm{obs}}$ extend from 0 to 1 here. If photon escapes
the cluster the event looks the same as for BARE1. The above
restrictions on $x$--integration limits can be written
symbolically as follows:
\begin{equation}\label{3}
\int\limits_{x_c}^{1}\dd x\ +\ \Theta(\delta - |\vecc r|)
\int\limits_{0}^{x_c}\dd x \ =\  \int\limits_{0}^{1}\dd x\
-\ \Theta(|\vecc r| - \delta)\int\limits_{0}^{x_c}\dd x,
\qquad \vecc r = \frac{\vecc k}{\omega}
- \frac{\vecc q_1^{\bot}}{\varepsilon_1}\, ,
\end{equation}
where $\vecc k$ $(\vecc q_1^{\bot})$
and $\omega$ $(\varepsilon_1)$ are the transverse momentum and
the energy of the hard photon (electron).

As we saw on the example of  BARE1 event selection it is necessary
to distinguish the contributions into $\Sigma_1$ due to electron
and positron radiation:
\begin{equation}\label{4}
\Sigma_1 = \Sigma^{\gamma} + \Sigma_{\gamma}\, .
\end{equation}
According to~(\ref{3}) we have
\begin{equation}\label{5}
\Sigma^{\gamma} = \Sigma_i + \Sigma_f + \Sigma_i^{^{c}}
+ \Sigma_f^{^{c}} \ , \quad
\Sigma_{\gamma} = \widetilde\Sigma_i + \widetilde\Sigma_f +
\widetilde\Sigma_i^{^{c}} + \widetilde\Sigma_f^{^{c}} \ ,
\end{equation}
where index $i(f)$ labels the initial (final) state
and index $c$ points on a cluster--form dependence.

Quantities $\Sigma_i$ and $\widetilde\Sigma_i$ coincide with
the corresponding initial--state correction for BARE1
(see comments on Eq.(1)). For $\Sigma_f$ and
$\widetilde\Sigma_f$ we can use the form of differential
cross--section suitable for inclusive event selection
with extended $x$-integration limits:
\begin{eqnarray}
\Sigma_f &=& \frac{\alpha}{2\pi}\int\limits_{\rho_2^2}^{\rho_4^2}
\frac{\dd z}{z^2}\biggl[-\frac{1}{2} + \int\limits_{0}^{1}(1-x +
\frac{1+x^2}{1-x}L_1) \dd x\biggr] \ , \nonumber \\
\widetilde\Sigma_f &=& \frac{\alpha}{2\pi}\int\limits_{1}^{\rho_3^2}
 \frac{\dd z}{z^2}\biggl[-\frac{1}{2}\Delta_{42} + \int\limits_{0}^{1}
\biggl((1-x + \frac{1+x^2}{1-x}\widetilde L_1)\Delta_{42}
\label{6}
+ \frac{1+x^2}{1-x}(\overline{\Theta}_4 - \Theta_2)\widetilde L_4\biggr)
\dd x\biggr] \, .
\end{eqnarray}

In order to find the additional contributions into $\Sigma_1$
which depend on the cluster form it is enough to use the
simplified differential cross--section of single photon radiation,
neglecting electron mass, and taking into account the restrictions
$ |\vecc r| < \delta$ (for the initial state) and
$ |\vecc r| > \delta$ (for the final state). The contribution due
to photon emission by the initial--state electron can be written
as follows:
\begin{equation}\label{7}
\Sigma_i^{^{c}} = \frac{\alpha}{2\pi}\int\limits_{
0}^{x_c}\frac{1+x^2}{1-x} \dd x\int\frac{\dd z}{z^2}\int
\dd z_1\;\Psi\,\Phi(z_1,z;x,\lambda) ,
\qquad \lambda = \frac{\delta} {\theta_1} \, ,
\end{equation}
where $\Psi$ defines the integration limits over
$z$ (in the square brackets) and over $z_1$ (in the parenthesis):
\begin{eqnarray*}
\Psi &=& [a^2,a_0^2](x^2z_+,x^2) + [b^2,a^2](x^2z_+,x^2z_-)
+ [b_0^2,b^2] (x^2\rho_3^2,x^2z_-),\quad
a_0 = \rho_2,\quad b_0 = \rho_4, \\
a &=& \mbox{max}(\rho_2, 1+\lambda(1-x)), \quad
b  =  \mbox{min}(\rho_4, \rho_3 - \lambda(1-x)), \quad
z_{\pm} = (\sqrt{z}\pm  \lambda(1-x)^2) \ ,
\end{eqnarray*}
and function $\Phi$ is defined below:
\begin{eqnarray}
\Phi(z_1,z;x,\lambda) &=& \frac{2}{\pi}\biggl(\frac{1}{z_1-xz}
+ \frac{1}{z-z_1}\biggr)\arctan
\biggl[\frac{(z-z_1)Q}{(\sqrt{z}-\sqrt{z_1})^2}\biggr],
\nonumber \\ \label{8}
Q &=& \sqrt{\frac{\lambda^2x^2(1-x)^2
- (\sqrt{z_1} - x\sqrt{z})^2}{(\sqrt{z_1}
+ x\sqrt{z})^2 - \lambda^2x^2(1-x)^2}} \ .
\end{eqnarray}

The cluster--dependent contribution due to photon emission
by the final--state electron reads
\begin{eqnarray*}
\Sigma_f^{^{c}} &=& \frac{\alpha}{2\pi}\int\limits_{
0}^{x_c}\frac{1+x^2}{1-x} \dd x\biggl[\int\frac{\dd z}{z^2}
\int \dd z_1\Psi F(z_1,z;x,\lambda)
+ \int\limits_{a_0^2}^{b^2}\frac{\dd z}{z^2}
\biggl(\ln\left|\frac{x\rho_3^2-z}{\rho_3^2-z}\right| + l_+ \biggr) \\
&+& \int\limits_{a^2}^{b_0^2}\frac{\dd z}{z^2}
\biggl(\ln\left|\frac{\rho_1^2 x - z}{\rho_1^2-z}\right|
+ l_-\biggr)\biggr], \label{9}
\qquad l_{\pm} = \ln\frac{\lambda(2\sqrt{z}
\mp\lambda(1-x))}{z\pm 2x\lambda\sqrt{z}-
\lambda^2x(1-x)}\, , \\
F &=& \frac{2}{\pi}\biggl(\frac{1}{z_1-xz}
- \frac{1}{z_1-x^2z}\biggr)\arctan
\biggl[\frac{(\sqrt{z_1}-x\sqrt{z})^2}{(z_1-x^2z)Q}\biggr].
\end{eqnarray*}
To obtain $\widetilde\Sigma_i^{^{c}}$ it is enough to substitute
in the expression for $\Psi$ the parameters  $a,\, b,\, a_0$
and $b_0$ by $\tilde a,\tilde b, \tilde a_0 $ and $\tilde b_0$
respectively,
\begin{eqnarray*}
\tilde a &=& \rho_2 + \lambda(1-x),\qquad
\tilde a_0 = \mbox{max}(1,\rho_2 - \lambda (1-x)), \\
\tilde b &=& \rho_4 - \lambda(1-x),\qquad \tilde b_0
= \mbox{min}(\rho_4 + \lambda(1-x),\rho_3).
\end{eqnarray*}

Finally, the cluster--dependent contribution due to photon
emission by the final--state positron can be written as follows:
\begin{eqnarray}
\widetilde\Sigma_f^{^{c}} &=& \frac{\alpha}{2\pi}\int\limits_{0}^{x_c}
\frac{1+x^2}{1-x}\dd x\int\limits_{1}^{\rho_3^2}
\frac{\dd z}{z^2}[\Theta(\tilde a_0^2 - z)
- \Theta(z - \tilde b_0^2)]\widetilde L_4
\nonumber \\ \label{10}
&+& \Sigma_f^{^{c}} (a,b,a_0,b_0 \rightarrow
\tilde a,\tilde b,\tilde a_0,\tilde
b_0;\ \ \rho_3, \rho_1\rightarrow\rho_4,\rho_2) \, .
\end{eqnarray}

The CALO2 event selection differs from the CALO1 one by the form
of the cluster (see [2]). Only cluster--dependent contributions
into $\Sigma_1$ will be changed here. Analytical formulae are
very cumbersome, and we give the result for symmetrical
wide--wide case only $(\Sigma^{\gamma} = \Sigma_{\gamma})$ but
our numerical calculations include wide--narrow angular
acceptance too.
\begin{eqnarray}
\Sigma_i^{^{c}} &=& \frac{\alpha}{2\pi}\int\limits_{0}^{x_c}
\frac{1+x^2}{1-x} \dd x\int\frac{\dd z}{z^2}\int
\dd z_1\frac{2}{\pi}\biggl(\frac{1}{z_1-xz} + \frac{1} {z-z_1}\biggr)
\label{11}
[\Psi_1\Phi_1 + \Psi_2\Phi_2 + \Psi_3\Phi_3] , \\ \nonumber
\Phi_1 &=& \arctan(Q_i^{(-)}) - \arctan(\eta), \qquad
\Phi_2 = \arctan(\eta^{-1}), \qquad
\Phi_3 = \arctan\left(\frac{1}{Q_i^{(+)}}\right), \\ \nonumber
\eta &=& r_i\cot\left(\frac{\Phi-\delta}{2}\right), \qquad
r_i = \frac{(\sqrt{z}-\sqrt{z_1})^2}{z-z_1}\, ,\\ \nonumber
Q_i^{(\pm)} &=& r_i\sqrt{\frac{x^2(\sqrt{z}+\sqrt{z_1})^2-(1-x)^2
(\sqrt{z_1}\pm x\bar\lambda)^2}{(1-x)^2(\sqrt{z_1}\pm x\bar\lambda)^2
- x^2(\sqrt{z}-\sqrt{z_1})^2}} \, , \\ \nonumber
\Psi_1 &=& [z_3^{(-)},1](x^2J_+^2,x^2z_+) +
[(\rho_3 - (1-x)\bar\lambda)^2, z_3^{(-)}](x^2\rho_3^2,x^2z_+), \\ \nonumber
\Psi_2 &=& [z_1^{(+)},1](x^2z_+,x^2) + [(\rho_3 - (1-x)\bar\lambda)^2,
z_1^{(+)}](x^2z_+,x^2J_-^2) \\ \nonumber
&+& [\rho_3^2 ,(\rho_3 - (1-x)\bar\lambda)^2](x^2\rho_3^2,x^2J_-^2),
\\ \nonumber
\Psi_3 &=& [z_1^{(+)},(1+(1-x)\bar\lambda)^2](x^2J_+^2,x^2) + [\rho_3^2 ,
(1+(1-x)\bar\lambda)^2](x^2J_-^2,x^2z_-).
\end{eqnarray}

We present also the corresponding formula for photon emission
by the final--state electron:
\begin{eqnarray}
\Sigma_f^{^{c}} &=& \frac{\alpha}{2\pi}\int\limits_{0}^{x_c}
\frac{1+x^2}{1-x} \dd x\;\biggl[\int\frac{\dd z}{z^2}\int\dd z_1\;
\frac{2}{\pi}\biggl(\frac{1}{z_1-xz} - \frac{1}{z_1-x^2z}\biggr)
[\Psi_1F_1 + \bar\Psi_2F_2 + \Psi_3F_3] \nonumber \\ \label{13}
&+& \int\limits_{1}^{z_3^{(-)}}
\frac{\dd z}{z^2}\ln\left|\frac{(x\rho_3^2-z)(J_+^2-z)}
{(\rho_3^2-z)(xJ_+^2-z)}\right|
+ \int\limits_{(1+(1-x)\bar\lambda)^2}^{\rho_3^2}\frac{\dd z}{z^2}
\biggl(\ln\left|\frac{x - z}{1-z}\right| + \bar l_-\biggr) \biggr],
\\ \nonumber
F_1 &=& \arctan\left(\frac{1}{Q_f^{(-)}}\right), \quad
F_2=\arctan(\zeta), \quad
F_3 = \arctan(Q_f^{(+)}), \quad
\zeta = r_f\cot\left(\frac{\Phi-\delta}{2}\right), \\ \nonumber
r_f &=&\frac{(\sqrt{z_1}-x\sqrt{z})^2}{z_1-x^2z}\, , \quad
\bar l_- = l_-(\lambda \rightarrow \bar\lambda), \quad
Q_f^{(\pm)} = \frac{r_f}{r_i}Q_i^{(\pm)}\, ,\quad
\sin(\delta) = \sqrt\frac{z_1}{z}\sin(\Phi), \\ \nonumber
\bar\Psi_2 &=& [z_1^{(+)},1](x^2J_+^2,x^2) + [z_3^{(-)},z_1^{(+)}]
(x^2J_+^2,x^2J_-^2)
+ [\rho_3^2 ,z_3^{(-)}] (x^2\rho_3^2,x^2J_-^2).
\end{eqnarray}

Angle $\Phi$ and quantity $\bar\lambda$, which enter into Eq.(12),
define the form and the size of CALO2 cluster. Namely~[2]
\begin{eqnarray*}
\Phi = \frac{3\pi}{32}\, , \qquad
\bar\lambda = \frac{\theta_0}{\theta_1}\, ,
\qquad \theta_0 = \frac{0.051}{16} \, .
\end{eqnarray*}

Finally, functions $J_{(\pm)}$ and $z_i^{(\pm)}$ are
defined as follows:
\begin{eqnarray*}
J_{(\pm)} &=& \frac{1}{\beta}\biggl[\sqrt{z\beta
- x^2(1-x)^2\bar\lambda^2
\sin^2\Phi} \pm (1-x)\bar\lambda(1-2x\sin^2\frac{\Phi}{2})\biggr] \ , \\
\beta &=& 1-4x(1-x)\sin^2\frac{\Phi}{2}, \quad
z_i^{(\pm)} = (\rho_i \pm (1-x)\bar\lambda)^2
- 4x(1-x)\rho_i(\rho_i \pm\bar\lambda)\sin^2\frac{\Phi}{2} \, .
\end{eqnarray*}

\section{Second and third order corrections }

In this Section we give the analytical form of the leading
second and third order corrections to SABS cross--section
suitable for both, inclusive and calorimeter, event selections.
The contribution connected with pair production in the singlet
channel is negligible for LEP1 conditions, while the one
in the non--singlet channel can be taken into account by means of
effective QED coupling~[6]. Therefore, we will consider here
the photonic corrections only.

The second order correction can be presented in the form
\begin{equation}\label{15}
\Sigma_2 = \Sigma^{\gamma\gamma} + \Sigma_{\gamma\gamma}
         + \Sigma_{\gamma}^{\gamma} \, .
\end{equation}
The first term in r.h.s. of Eq.(\ref{15}) is responsible for
two--photon (real and virtual) emission by the electron.
The second one describes two--photon emission by the positron.
And the third one considers the situation when both the electron
and the positron radiate.

The leading contributions in the case of inclusive event selection read
\begin{eqnarray}
\Sigma^{\gamma\gamma L} &=& \frac{\alpha^2}{4\pi^2}
\int\limits_{\rho_2^2}^{\rho_4^2}
\frac{\dd z}{z^2}L^2\int\limits_{x_c}^{1}\dd x\;
\biggl[\frac{1}{2}(1+\Theta_3^{(x)})P^{(2)}(x)
+ \int\limits_{x}^{1}\frac{\dd t}{t}P^{(1)}(t)P^{(1)}
\biggl(\frac{x}{t}\biggr)\Theta_3^{(t)}\biggr], \nonumber \\ \nonumber
\Sigma_{\gamma\gamma}^L &=& \frac{\alpha^2}{4\pi^2}
\int\limits_{1}^{\rho_3^2}\frac{\dd z}{z^2}L^2
\int\limits_{x_c}^{1}\dd x\;\biggl[\frac{1}{2}(\Delta_{42}
+\Delta_{42}^{(x)})P^{(2)}(x) + \int\limits_{x}^{1}
\frac{\dd t}{t}P^{(1)}(t)P^{(1)}
\biggl(\frac{x}{t}\biggr)\Delta_{42}^{(t)}\biggr], \\
\Sigma_{\gamma}^{\gamma L} &=& \frac{\alpha^2}{4\pi^2}
\int\limits_{0}^{\infty}\frac{\dd z}{z^2}L^2
\int\limits_{x_c}^{1}\dd x_1\int\limits_{\frac{x_c}{x_1}}^{1}
\dd x_2\; P^{(1)}(x_1)P^{(1)}(x_2)                       \label{16}
(\Delta_{31} +\Delta_{31}^{(x_1)}) (\Delta_{42}
+ \Delta_{42}^{(x_2)}).
\end{eqnarray}
where
\begin{eqnarray*}
P^{(2)}(x) = \int\limits_{x}^{1}\frac{\dd t}{t}P^{(1)}(t)
P^{(1)}\left(\frac{x}{t}\right) , \qquad
\int\limits_{0}^{1}P^{(2)}(x)\dd x\; = 0.
\end{eqnarray*}

The contribution of initial--(final--)state radiation for
$\Sigma^{\gamma\gamma L}$ and $\Sigma_{\gamma\gamma }^{L}$
are accompanied with $x$-dependent ($x$--independent)
$\Theta$--functions and $x_1,x_2$--dependent
($x_1,x_2$--independent) ones for $\Sigma_{\gamma}^{\gamma L}$.
The terms with additional integration over $t$--variable
describe the simultaneous radiation of one photon from the
initial state and the other from final state
(initial--final--state radiation).

In the case of calorimeter event selection we have to take
in the r.h.s. of Eqs.(14) the terms corresponding to
initial--state radiation only. The elimination of
final--state one exhibits itself by means the last equality.
As concerns the contribution due to initial--final--state
radiation it may be understood as follows.

In fact $t$--variable in Eqs.(14) is the energy fraction
carried out by both the final--state radiated  photon and
the final--state electron (or positron). Just this value
defines the cluster energy for calorimeter event selection.
The $x$--variable which is the energy fraction of the final
electron by definition can change here from 0 up to $t$.
That is why initial--final--state radiation of the electron
for calorimeter event selection will be proportional to
\begin{equation}\label{17}
\int\limits_{x_c}^{1}\dd t\int\limits_{0}^{t}\frac{\dd x}{t}
P^{(1)}(t)P^{(1)} \left(\frac{x}{t}\right)\Theta_3^{(t)}
= \int\limits_{x_c}^{1}\dd t\; P^{(1)}(t)
\Theta_3^{(t)}\int\limits_{0}^{1}P^{(1)}(y)\dd y\; = 0.
\end{equation}
The same is valid of course for the corresponding part of
positron radiation.

Thus, we see that it is enough to have only single
final--state radiated photon to eliminate the leading
contribution due to initial--final--state radiation.
This conclusion reflects the essence of a reduced Lee--Nauenberg
theorem~[12] and is valid for all higher order corrections.

The leading third order correction reads
\begin{eqnarray} \label{18}
\Sigma_3^{^{L}} &=& \left(\frac{\alpha}{2\pi}\right)^3
\int\limits_{0}^{\infty}\frac{\dd z}{z^2}
L^{^{3}}\int\limits_{x_c}^{1}\biggl(Z_1
+ \int\limits_{x_c/x}^{1}Z_2\dd x_1
\biggr)\dd x\; , \\ \nonumber
Z_1 &=& \frac{1}{6}(2\Delta_{42} + \Delta_{42}^{(x)}\Delta_{31}
+ \Delta_{31}^{(x)}\Delta_{42})P^{(3)}(x) \\ \nonumber
&+& \frac{1}{2}\int\limits_{x}^{1}\frac{\dd t}{t}
(\Delta_{42}^{(t)}\Delta_{31} + \Delta_{31}^{(t)}\Delta_{42})
\biggl[P^{(1)}(t)P^{(2)}\left(\frac{x}{t}\right)
+ P^{(2)}(t)P^{(1)}\left(\frac{x}{t}\right)\biggr],
\\ \nonumber
Z_2 &=& \frac{1}{2}\bigl[(\Delta_{31}+\Delta_{31}^{(x)})
(\Delta_{42}+\Delta_{42}^{(x_1)})
+ (\Delta_{31}+\Delta_{31}^{(x_1)})(\Delta_{42}+\Delta_{42}^{(x)})\bigr]
P^{(1)}(x)P^{(2)}(x_1) \\ \nonumber
&+& P^{(1)}(x)\int\limits_{x_1}^{1}\frac{\dd t}{t}
[\Delta_{31}^{(t)}(\Delta_{42}+\Delta_{42}^{(x)})
+ \Delta_{42}^{(t)}(\Delta_{31}+\Delta_{31}^{(x)})]
P^{(1)}(t)P^{(1)}\left(\frac{x_1}{t}\right), \\ \nonumber
P^{(3)} &=& \int\limits_{x}^{1}\frac{\dd t}{t}P^{(1)}(t)\,
          P^{(2)}\left(\frac{x}{t}\right).
\end{eqnarray}

For calorimeter event selection it is needed to take
\begin{eqnarray*}
Z_1 = \frac{1}{6}(\Delta_{42}^{(x)}\Delta_{31}+\Delta_{31}^{(x)}
\Delta_{42})P^{(3)}(x), \quad
Z_2 = \frac{1}{2}(\Delta_{31}^{(x)}\Delta_{42}^{(x_1)}+\Delta_{31}
^{(x_1)}\Delta_{42}^{(x)})P^{(1)}(x)P^{(2)}(x_1).
\end{eqnarray*}
Note that in this case the leading second and third
order contributions have a universal character and
do not depend on cluster form. Thus, they are suitable
for both,  CALO1 and  CALO2.

Quantity $Z_1$ describes the situation when only one fermion
(electron or positron) radiate (one--side emission), while
$Z_2$ is responsible for simultaneous radiation of the electron
and the positron (opposite--side emission).

The formulae for leading second and third order contributions
are written in the form with  different $\Theta$--functions
under integral sign. One can eliminate these $\Theta$--functions
using such kind relations as, for example
\begin{equation}\label{19}
\int\Theta_4\bar\Theta_4^{(x)}\bar\Theta_3^{(x_1)}\dd z\;\dd x\;
\dd x_1\; = \int\limits_{x_c\rho_3}^{\rho_4^2}\dd z
\int\limits_{x_c\rho_3/\sqrt{z}}^{\sqrt{z}/\rho_4}\dd x
\int\limits_{x_c/x}^{\sqrt{z}/\rho_3}\dd x_1\, .
\end{equation}
It is needed to keep in mind only that every integral has to be
equal to zero if the lower limit of $z$-integration becomes more
than the upper one. The last statement is valid for the first
order correction too.

\section{Numerical results}

In our calculations we restrict ourselves  with pure QED
corrections supposing $Z$--exchange, vacuum polarization
and up--down interference are {\em switched off.\/} As
shown in papers by W.~Beenakker and B.~Pietrzyk~[7],
a sufficiently accurate luminosity determination requires the
full Born plus complete order $\alpha$ corrected cross--section.
Nevertheless our numerical results can be used for comparisons
and cross--checks with numerous Monte Carlo and semi--analytical
computations~[2].

We performed the numerical calculations for the beam energy
$\sqrt{s}/2 = 46.15$~GeV
and the following sets of limiting angles of circular detectors:
\begin{eqnarray*}
&& \mbox{i)\  BARE1, CALO1:}\ \theta_1=0.024,\ \theta_3=0.058,\ \
\theta_2=\theta_1+h,\ \theta_4=\theta_3-h, \ h=\frac{0.017}{8} \ ; \\
&& \mbox{ii)\ CALO2:}\ \theta_1=0.024+h,\ \theta_3=0.058-h,\ \
\theta_2=\theta_1+h,\ \theta_4=\theta_3-3h \, .
\end{eqnarray*}

  The Born cross--section
\begin{equation}\label{20}
\sigma_B =  \frac{4\pi\alpha^2}{Q_1^2}\int
\frac{\dd z}{z^2}\left(1 - \frac{z\theta_1^2}{2} \right)
\end{equation}
(limits of integration are $(\rho_4^2, \rho_2^2)$ for NN and WN
angular acceptances and $(\rho_3^2, 1)$ for WW one) equals to
\begin{eqnarray*}
&& 175.922\ \mbox{nb  \ \ ----- \ \   BARE1,\ CALO1 \ WW}, \\
&& 139.971\ \mbox{nb \ \ ----- \ \   BARE1,\ CALO1 \ NN;\ CALO2 \ WW }, \\
&& 103.299\ \mbox{nb  \ \ ----- \ \   CALO2 \ NN }.
\end{eqnarray*}

Formula~(\ref{20}) takes into account the contributions of
scattering diagram and interference of scattering and
annihilation ones. The contribution of pure annihilation
diagram is proportional to $\theta_1^4$. It is negligible
even at the born level. When calculating the QED corrections to
cross--section~(\ref{20}) we ignore systematically
the terms proportional $\theta_1^2$. Terms of this kind
have double logarithmic asymptotic behavior [13] and are equal
parametrically to $ (\alpha|t|)\ln^2(|t|/s)/(\pi s)$,
what is about 0.1 $per \ mille$
as compared with unit for LEP1 conditions.

\begin{table}
\caption{The results of our analytical calculations
for the SABS cross--section }
\begin{tabular}{rrrrrrrrrr}  \hline
        & \multicolumn{3}{c}{BARE1} & \multicolumn{3}{c}{CALO1} &
\multicolumn{3}{c}{CALO2} \\ \hline
$x_c$   & WW & NN & WN & WW & NN & WN & WW & NN & WN \\ \hline
    & \multicolumn{9}{c}{$\sigma$ (nb) with ${\cal O}(\alpha^1)$
corrections} \\  \hline
0.1 & 166.008 & 130.813 & 134.504 & 166.285 & 131.032 & 134.270 &
130.997 & 94.666 & 98.354  \\
0.3 & 164.702 & 129.797 & 133.416 & 166.006 & 130.833 & 134.036
& 130.705 & 94.491  & 98.127   \\
0.5 & 162.203 & 128.001 & 131.428 & 165.244 & 130.416 & 133.466
& 130.141 & 94.177  & 97.720   \\
0.7 & 155.390 & 122.922 & 125.809 & 161.749 & 128.044 & 130.542
& 127.491 & 92.981  & 95.874   \\
0.9 & 134.334 & 106.478 & 107.945 & 149.866 & 118.822 & 120.038
& 117.491 & 86.303  & 87.696   \\  \hline
    & \multicolumn{9}{c}{$\sigma$ (nb) with ${\cal O}(\alpha^1)$
 and ${\cal O}(\alpha^2)$ photonic corrections} \\ \hline
0.1 & 166.958 & 131.674 & 134.808 & 167.073 &131.740 & 134.572 &
131.705 & 95.334 & 98.609  \\
0.3 & 165.447 & 130.534 & 133.583 & 166.686 & 131.467 & 134.231 &
131.339 & 95.118 & 98.314 \\
0.5 & 162.574 & 128.474 & 131.127 & 165.718 & 130.903 & 133.471 &
130.628 & 94.731 & 97.793 \\
0.7 & 155.597 & 123.206 & 125.255 & 162.042 & 128.361 & 130.378 &
127.808 & 93.377 & 95.782 \\
0.9 & 137.153 & 108.820 & 109.677 & 150.732 & 119.560 & 120.411 &
118.229 & 86.931 & 87.961 \\  \hline
   &  \multicolumn{9}{c}{absolute values of the ${\cal O}(\alpha^2)$ pair
production correction (nb)} \\  \hline
0.1 & 0.007 & $-$0.004 & 0.015 & $-$0.046 & $-$0.045 & $-$0.024 & $-$0.045 & $-$0.047
& $-$0.024 \\
0.3 & $-$0.033 & $-$0.033 & $-$0.020 & $-$0.046 & $-$0.045 & $-$0.024 & $-$0.045
& $-$0.047 & $-$0.024 \\
0.5 & $-$0.058 & $-$0.050 & $-$0.041 & $-$0.048 & $-$0.046 & $-$0.025 & $-$0.046 &
$-$0.047 & $-$0.024 \\
0.7 & $-$0.090 & $-$0.074 & $-$0.069 & $-$0.069 & $-$0.059 & $-$0.042 & $-$0.059 &
$-$0.051 & $-$0.036 \\
0.9 & $-$0.142 & $-$0.115 & $-$0.115 & $-$0.137 & $-$0.111 & $-$0.102 & $-$0.111 &
$-$0.085 & $-$0.075 \\   \hline
   &  \multicolumn{9}{c}{absolute value of the ${\cal O}(\alpha^3)$ leading
correction (nb)} \\  \hline
0.1 & $-$0.055 & $-$0.047 & $-$0.006 & $-$0.041 & $-$0.036 & $-$0.002 & $-$0.036 &
$-$0.034 & $-$0.001  \\
0.3 & $-$0.065 & $-$0.053 & $-$0.018 & $-$0.046 & $-$0.040 & $-$0.007 & $-$0.040 &
$-$0.037 & $-$0.003 \\
0.5 & $-$0.038 & $-$0.040 & 0.004 & $-$0.044 & $-$0.039 & $-$0.006 & $-$0.039 &
$-$0.037 & $-$0.005 \\
0.7 & 0.089 & 0.058 & 0.124 & $-$0.023 & $-$0.022 & 0.012 & $-$0.022 & $-$0.027 &
0.008 \\
0.9 & 0.291 & 0.220 & 0.331 & 0.021 & 0.013 & 0.049 & 0.013 & 0.002 & 0.038
\\  \hline
\end{tabular}
\end{table}

\begin{table}
\caption{BHLUMI results for the small--angle
Bhabha cross--section (in nb)}
\begin{tabular}{rrrrcrrrr} \hline
   & \multicolumn{3}{c}{${\cal O}(\alpha^1)$} &
\multicolumn{5}{c}{beyond the first order}  \\ \hline
$x_c$   & BARE1 & CALO1 & CALO2 & BARE1
& CALO1 & \multicolumn{3}{c}{CALO2} \\ \hline
0.1 & 166.046 & 166.329 & 131.032 & 166.892~(988)~[0.008] & 167.203
& 131.835 & 95.458 & 98.834 \\
0.3 & 164.740 & 166.049 & 130.739 & 165.374~(471)~[0.014] & 166.795
& 131.450 & 95.233 & 98.539 \\
0.5 & 162.241 & 165.287 & 130.176 & 162.530~(594)~[0.018] & 165.830
& 130.727 & 94.841 & 98.020 \\
0.7 & 155.431 & 161.794 & 127.528 & 155.688~(620)~[0.018] & 162.237
& 127.969 & 93.520 & 96.054 \\
0.9 & 134.390 & 149.925 & 117.541 & 137.342~(191)~[0.018] & 151.270
& 118.792 & 87.359 & 88.554 \\ \hline
\end{tabular}
\end{table}

The results of our calculations of QED correction with the
{\em switched off\/} vacuum polarization are shown in the
Table~1. The centre--of--mass energy is $\sqrt{s} = 92.3$~GeV.
The second order correction
in the case BARE1 contains both leading and next--to--leading contributions.
In the rest cases the higher order corrections are take in the leading
approximation.
For a comparison we give in Table~2 also the corresponding numbers
derived by the help of Monte Carlo generator  BHLUMI~[2].
Parameters are the same as for Table~1.
The results of the non--exponentiated version of BHLUMI for BARE1
differs from the exponentiated ones by three digits after decimal
point, which are given in parenthesis.
The numbers in square brackets
are absolute difference (in nb) between our second order photonic
correction and the one of the non--exponentiated  BHLUMI version.
BHLUMI numbers beyond the
first order for CALO2 case correspond to WW, NN and WN angular
acceptances respectively; and the rest is for WW case.
Beyond the first order all  BHLUMI
numbers, except the ones in parenthesis for BARE1, correspond to the
version based on the Yennie--Frautchi--Suura exponentiation.

On the level of the first order correction BHLUMI numbers exceed
our ones approximately on $0.3$ {\em per mille\/} for all variants
of event selection. We think that this is due to the difference
in our approaches: BHLUMI computes the first order correction~[14]
according to complete ${\cal O}(\alpha)$ formulae,
while we take into account only $t$--channel graphs as discussed above.

To be consequent we have to compare our results due to second order
photonic contribution with  BHLUMI ones which belongs to the
non--exponentiated version only. These are the numbers into the
parenthesis for BARE1 (three figure after point in the cross--section).
To compare it needs to remove the difference due to the first
order contribution. After this we find
that our second order photonic correction which includes leading and
next--to--leading contributions exceeds a little bit the  BHLUMI one and
conclude about very expressive agreement in the case of  BARE1~~WW.

As concerns calorimeter event selection we have not explicit
calculation of the second order next--to--leading contribution.
That is why now we can speak about the first order correction only.
As one can see from the Tables the agreement
of our numbers for  WW variant of  CALO1 and  CALO2 with
BHLUMI ones is on the same level as for BARE1~~WW one.

\ack
Authors thank S.~Jadach, G.~Montagna, B.~Pietrzyk and
B.~Ward for fruitful discussions. This work was supported
in part by the INTAS Grant 93--1867.

\end{document}